# Voltage Controlled Energy Efficient Domain Wall Synapses with Stochastic Distribution of Quantized Weights in the Presence of Thermal Noise and Edge Roughness

Walid Al Misba[1], Tahmid Kaisar[1], Dhritiman Bhattacharya[1], Jayasimha Atulasimha[1,2]

[1]Dept. of Mech. & Nuc. Engineering, Virginia Commonwealth University, Richmond, VA, US

[2]Dept. of Elec. & Comp. Engineering, Virginia Commonwealth University, Richmond, VA, US

**Abstract:**

We propose energy efficient strain control of domain wall (DW) in a perpendicularly magnetized nanoscale racetrack delineated on a piezoelectric substrate that can implement multi state synapse to be utilized in neuromorphic computing platforms. In conjunction with SOT from to a current flowing in the heavy metal layer, strain is generated by applying a voltage across the piezoelectric. Such a strain is mechanically transferred to the racetrack and modulates the Perpendicular Magnetic Anisotropy (PMA). When different voltages are applied (i.e. different strains are generated), it can translate the DW to different distances for the same current which implements different synaptic weights. We have shown using micromagnetic simulations that 5-state and 3-state synapse can be implemented in a racetrack that is modeled with natural edge roughness and room temperature thermal noise. Such strain-controlled synapse has an energy consumption of few fJs and could thus be very attractive to implement energy-efficient quantized neural networks, which has been shown recently to achieve near equivalent classification accuracy to the full-precision neural networks.

**Introduction:**

Neuromorphic computing outperforms traditional von-Neumann type processors in data-intensive classification tasks. Moreover, their in-memory computing architecture can reduce energy dissipation [1] required to shuttle data back and forth between processor and memory unit in traditional computing architectures. Examples of hardware realization for neuromorphic computing include phase change random access memory (PCRAM) [2-4], resistive random-access memory (RRAM) [5,6] and STTRAM [7]. While the device variability is a persistent issue for all of the above-mentioned devices, recent work in fully connected artificial neural network (ANN) [8] shows equivalent accuracy to software-based training. Unfortunately, PCRAM and RRAM based devices consume energy on the order of a few pJs per synaptic weight alteration event [9]. Hence, the future IoTs and edge-devices where power is limited will necessitate alternate neuromorphic hardware that are energy efficient and enable real time programing of synaptic weights so the networks can be trained in-situ.

Recently, nanomagnet based synaptic devices has shown potential to be energy efficient compared to PCRAM and RRAM [9, 10, 11]. Among nanomagnet based neuromorphic devices, domain wall (DW) based MTJs are one of the most promising. To implement these devices, domain walls (DWs) are translated to different positions by externally applied magnetic field [12], an electric current that causes Spin-Orbit Torque (SOT) [13-15], Spin Transfer Torque (STT) [16-18] or a strain gradient [19-20]. Strain control of magnetization consumes ultra-low energy [21-27]. Hence, manipulation of DWs with strain can be utilized to implement energy efficient neuromorphic devices. Recently, strain-mediated control of DW has been reported [28, 29]. Strain gradient in conjunction with SOT or STT [10] has also been proposed to control DW position to implement energy efficient synaptic devices that can be programmed in real time.

In this work, we propose to utilize SOT to translate the DW in a realistic nanoscale racetrack modeled with edge roughness and thermal noise where the DW position is controlled by modulating the Perpendicular Magnetic Anisotropy (PMA) of the racetrack with the application of stress. Here, deterministic control of DW to realize different synaptic values is hard to achieve when different stress values are generated by applying voltage pulse of different amplitudes to the electrodes patterned on top of a piezoelectric. This is because, equilibrium DW positions are often stochastic in nature and with the presence of defects [30], local imperfections [31] and thermal noise [32] it could be very difficult to achieve deterministic control. Nevertheless, the DW can be arrested by providing trap sites such as curved shape [33] and notch or protrusion [34], which can act as a potential well or barrier. Moreover, edge roughness [35-36] can introduce pinning sites for DW motion. In this paper, we used edge roughness and obtain the statistical distribution of DW position from micromagnetic simulations which showed that the mean positions are different for different stress induced change of PMA for a fixed current induced SOT of a fixed "clock" time. Although the number of states (different DW positions) attained are limited and there are overlaps between the states, such a DW based racetrack as synapse is particularly attractive to implement quantized deep learning neural networks [37-39] as these networks have been shown to reach accuracy very close to the infinite states network. Moreover, the overlap between states could be addressed during the training stage of a learning network. Infact, the overlap between states would be useful in generating stochastic weight distribution for training the network in-situ as training with stochastically determined weights rather than deterministic ones can potentially increase the classification accuracies for some data sets [37].

**Device architecture and simulation:**

The proposed device structure is illustrated in Fig. 1a. The stack consists of a heavy metal layer and an MTJ containing the nanoscale racetrack as free layer, along with the tunnel barrier and the hard layer. Such a stack is patterned on top of a piezoelectric substrate. We consider Pt/CoFe (soft or free racetrack layer)/MgO/CoFe (hard or fixed layer) as our stack materials where the heavy metal layer Pt will create perpendicular anisotropy and strong DMI at Pt/CoFe interface, which is known to favor the chiral Neel DWs [40]. We propose to arrest the DWs at different positions in the free layer of the MTJ, which will modify the resistance value of the MTJ stack. Thus, different synaptic weights, which define the strength between the neurons can be determined from the DW positions.

To arrest the DW at various positions we apply different amplitude stress in combination to a fixed amplitude and fixed duration SOT pulse. When a voltage is applied between the electrodes on top and bottom of the piezo-substrate as shown in Fig. 1b, mechanical strain is generated. This strain is then transferred to the racetrack and consequently modulates the perpendicular anisotropy due to magnetoelastic interaction. In combination with stress, we apply a current pulse in the adjacent Pt layer to exert SOT, which moves the DW through the nanowire racetrack to the other end of the nanowire. If we reverse the direction of current in the heavy metal layer, it will reverse the direction of DW motion and reset it to the other end.

We have considered edge roughness that is present naturally in a nanowire due to lithographic imperfection. We assumed normal Gaussian distribution where the rms edge roughness was taken to be ~ 3 nm. The simulated nanowires have a length of 500 nm, maximum width of 50 nm and thickness of 1 nm. The magnetization dynamics in the presence of Spin Orbit Torque (SOT) is simulated in MUMAX3 [41] using the Landau–Lifshitz–Gilbert-Slonczewski equation:

$$(1 + \alpha^2)\frac{d\vec{m}}{dt} = -\gamma \vec{m} \times \vec{H}_{eff} - \alpha\gamma \left(\vec{m} \times \left(\vec{m} \times \vec{H}_{eff}\right)\right) - \beta\gamma(\varepsilon - \alpha\varepsilon')\left(\vec{m} \times (\vec{m}_P \times \vec{m})\right) \\ + \beta\gamma(\varepsilon' - \alpha\varepsilon)(\vec{m} \times \vec{m}_P) \quad (1)$$

$$\beta = \frac{\hbar J\theta}{\mu_0 e d M_s}, \quad \varepsilon = \frac{P\Lambda^2}{(\Lambda^2+1)+(\Lambda^2-1)(\vec{m}\cdot\vec{m}_p)} \tag{2}$$

We consider secondary spin torque parameter to be $\varepsilon' = \alpha\varepsilon$ and neglect the field like torque. The spin current direction is $\vec{m}_P = \vec{J}_x \times \vec{z}$, $J$ is the value of current flowing through the heavy metal layer, $\vec{J}_x$ is the unit vector defining the direction of current flow and $\vec{z}$ is the direction of inversion asymmetry. Here, $\theta$ is the spin Hall angle which is 0.1 for Pt [42], $\gamma$ is the gyromagnetic ratio, $\vec{m}$ is the unit magnetization vector, $M_s$ is the saturation magnetization, $\hbar$ is the reduced Planck constant, $\mu_0$ is the permeability of free space, e is the electron charge and $d$ is the thickness of the nanowire. To equate the Slonczewski toque with Spin orbit torque we assume spin polarization to be $P = 1$ and Slonczewski parameter to be $\Lambda = 1$. Here the effective field, $\vec{H}_{eff}$ accounts for the contribution from PMA, Heisenberg exchange interaction, Dzyaloshinskii–Moriya interaction (DMI), stress induced anisotropy and thermal noise.

The racetracks are discretized into 2 nm x 2 nm x 1 nm cells which are well within the ferromagnetic exchange length of $\sqrt{\frac{2A_{ex}}{\mu_0 M_{sat}^2}}$ =5.66 nm.

PMA induced effective field can be expressed as, $\vec{H}_{anis}$:

$$\vec{H}_{anis} = \frac{2K_u}{\mu_0 M_s}(\vec{u}\cdot\vec{m})\vec{u} \tag{3}$$

Where $K_u$ is the first order anisotropy constant and $\vec{u}$ represents the uniaxial anisotropy direction (i.e. perpendicular to plane).

If the electrodes patterned on top of the piezoelectric substrate have dimensions similar to the piezoelectric thickness and separated by one or two times the piezoelectric thickness, maximum stress is generated [43]. In such a scenario, when a positive (negative) voltage is applied in the top electrode pair, the area underneath the electrode become stretched (compressed) in the out of plane direction and compressed (stretched) in the in-plane direction. Compression (tension) in the in-plane direction underneath the electrode surface creates tension (compression) in the nanoscale racetrack patterned in between the top electrodes due to strain-displacement compatibility. We assumed our electrodes to be rectangular with width b=piezoelectric thickness and length L=racetrack length. This is similar to having (L/b) number of square electrodes of (b×b) dimensions and therefore one can assume this electrode configuration will produce similar amount of stress as mentioned in Ref [50]. Fig. 1b shows the strain formation in the nanoscale racetrack in such a scenario. Stress produced in the in-plane direction of the racetrack induces anisotropy field due to the magneto-elastic effect in the same direction and modulates the PMA or the anisotropy constant $K_u$. The effect of the stress is modeled by the modulating $K_u$ in the micromagnetic simulation. For simplicity, we did not consider the strain that can be produced in the in-plane direction of the racetrack which is orthogonal to that shown in the Fig. 1b. The parameters for the simulation are presented at table 1.

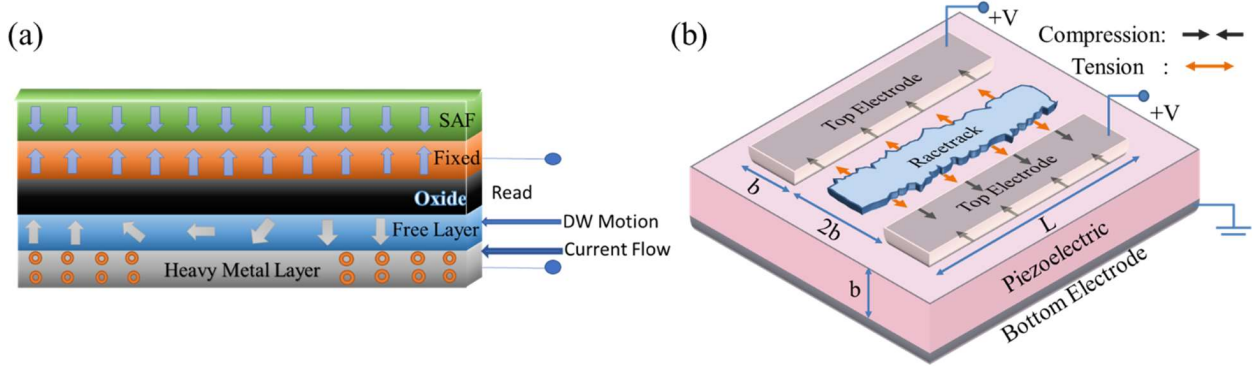

FIG. 1. **a.** Proposed device stack where the nanoscale racetrack is the free layer. DW in the racetrack moves when a current is applied to the heavy metal layer **b**. Stress generation mechanism in rough edge racetrack when a voltage is applied across the piezoelectric.

Table 1: Material parameters used for the CoFe soft layer in the Pt/ CoFe/MgO heterostructure from prior research works [44-46].

| Parameters | Values |
|---|---|
| DMI constant (D) | 0.001 Jm$^{-2}$ |
| Gilbert damping (α) | 0.015 |
| Saturation magnetization ($M_{sat}$) | $10^6$ Am$^{-1}$ |
| Exchange constant ($A_{ex}$) | $2\times10^{-11}$ Jm$^{-1}$ |
| Saturation magnetostriction ($\lambda_s$) | 250 ppm |
| Perpendicular Magnetic Anisotropy ($K_u$) | $7.5\times10^5$ Jm$^{-3}$ |

1. **Non-thermal statistics due to different edge roughness profiles in different racetracks**

For non-thermal simulations we have simulated the DW motion in 40 different racetracks with different edge roughness profile. The PMA of the racetracks is considered to be $7.5 \times 10^5$ J/m³. The PMA can be decreased or increased uniformly over the whole racetrack by applying a suitable voltage to the electrodes. The clocking SOT current is applied simultaneously with this voltage pulse. We have assumed that the DW is initialized to a pinning site located at one end of the racetrack. The SOT current translates the DW while the PMA modulation helps to drive the DW to different positions when clocked with SOT for a fixed time. This could be explained as following.

The critical depinning current density $J_C$ of the DW is related to the anisotropy coefficient $K_u$ of the racetrack. When $K_u$ is higher, the potential well of a pinning site becomes deep, so it requires high depinning current, $J_C$ to depin a DW siting in such a potential well or energy minima. On the contrary, lower $K_u$ is associated with a shallow potential well for the same pinning site hence requires lower threshold current to depin. Fig. 2a presents a sketch of an example racetrack where the DW is situated at a pinning site located near the right end of the racetrack and Fig. 2b plots the depinning current versus the anisotropy coefficient for that DW. From Fig. 2b we can see that critical depinning current $J_C$ is increased with the increase of anisotropy coefficient $K_u$.

The DW velocity at steady state can be expressed by the following [47,48]:

$$v = \frac{\pi}{2} \frac{\gamma \Delta H_{DMI}}{\sqrt{(1 + (\frac{J_D}{J - J_C})^2)}} \qquad (4)$$

$$J_D = \alpha J\, H_{DMI}/H_{SH} \qquad (5)$$

$$H_{DMI} = \frac{D}{\mu_0 \Delta M_s} \qquad (6)$$

$$H_{SH} = -\frac{\hbar J \theta}{2\mu_0 e d M_s} \qquad (7)$$

Where, $H_{DMI}$ [40] is the DMI field and $H_{SH}$ is the damping like spin hall effective field. Empirical critical current density $J_C$ is used to account for the pinning effect which is validated by fitting one dimensional DW model to the experimental data [48]. Here, $\Delta$ is the DW width which can be expressed as:

$$\Delta \sim \sqrt{\frac{A_{ex}}{K_u - \frac{1}{2}\mu_0 M_s^2}} \qquad (8)$$

As seen from Fig. 2b, the critical current density $J_C$ is high for higher Ku. As a result, for a higher Ku, for a fixed clocking SOT current $J > J_C$, the velocity becomes small as the denominator in Eq. 4 is large compared to the case of lower Ku for which the denominator is small (low critical current density $J_C$) and velocity is high. In addition, when Ku increases (decreases) the DW width $\Delta$ increases (decreases) which increases (decreases) $J_D$ in Eq. 5 and the denominator in Eq. 4, consequently the velocity decreases (increases).

The DW position for different anisotropy constant $K_u$ is shown in Fig. 2c for one rough edge racetrack where the SOT current of $24 \times 10^{10}\ A/m^2$ is applied for fixed 1.2 ns. The change in velocity with the change in $K_u$ is evident as the DW translates to different distances with the same SOT. After the withdrawal of the SOT and strain. the DW further moves at terminal velocity due to the momentum gained because of the SOT toque. The lower the anisotropy constant the higher the velocity gain and the higher the distance travelled by the DW after the withdrawal of SOT as can be seen for the case of $K_u = 7.3 \times 10^5\ J/m^3$. Notably, the DW for $K_u = 7.0 \times 10^5\ J/m^3$ also traveled same distance as $K_u = 7.3 \times 10^5\ J/m^3$ as the velocity difference after SOT withdrawal is small.

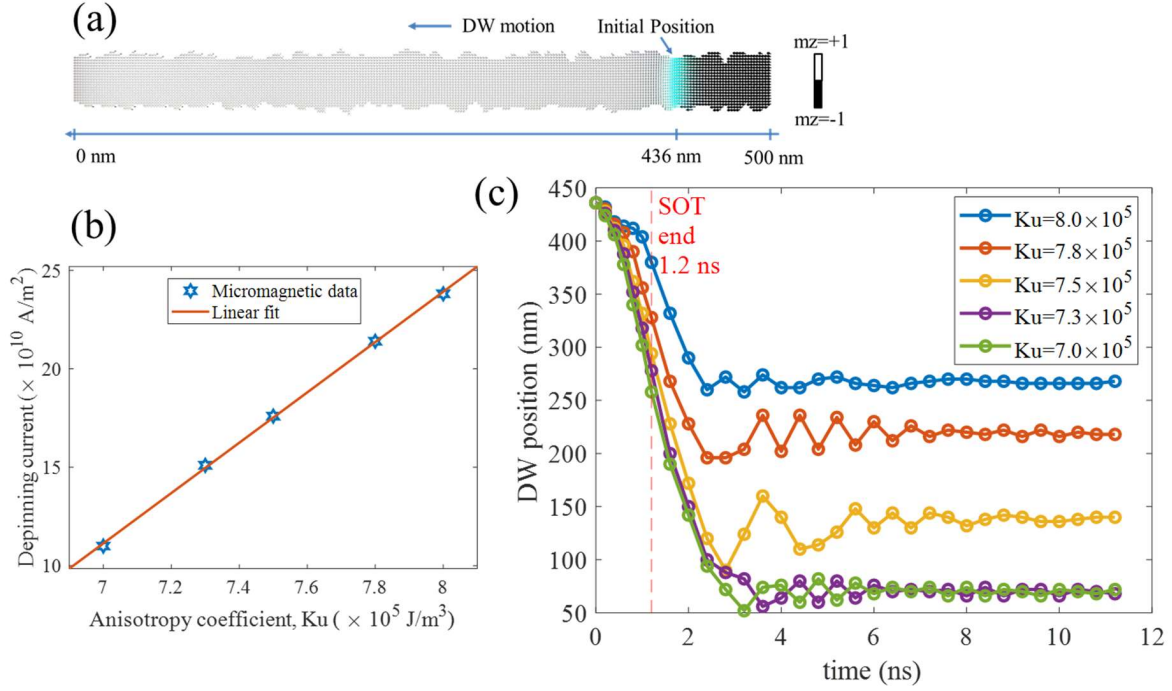

FIG. 2. **a.** Initial pinning position of the DW in a PMA rough edge racetrack **b.** dependence of the DW depinning current on the anisotropy coefficient when the DW in racetrack 2(a) is in the initial pinning position **c.** DW positions with time in racetrack 2(a) for a fixed duration and amplitude current pulse exerting SOT and different stresses (different Ku). The SOT and stress are withdrawn at 1.2 ns. For different stresses respective DWs travel different distances and get pinned to different locations.

We have simulated a total of 40 racetracks of ~ 3nm rms edge roughness where we varied anisotropy constant values $K_u$ to 8.0, 7.8, 7.5, 7.3 and 7.0 ($\times 10^5$) $J/m^3$ in each of these racetracks and applied SOT current of fixed amplitude $24 \times 10^9\ A/m^2$ for 1.2 ns. Each of the DW is initialized to a pinning site located near the right end of the racetrack. After the simultaneous withdrawal of the SOT and stress we wait for 10 ns to allow sufficient time for the DW to decelerate and get pinned to a specific position. The distribution of the final DW position for the 40 racetracks is shown in Fig. 3.

In Fig. 3 for each Ku value we also overlay a gaussian distribution with identical mean and standard deviation of the data used to create the bins. Although the final position distribution does not follow Gaussian distribution, we see that the mean final positions are different for different stress (Ku) values (Fig. 3f). The mean DW positions shift to the left of the racetracks as we decrease the PMA. The distribution of the DW final positions for a specific Ku could be attributed to the following. DWs in different racetracks are initialized from a pinning site having a longitudinal position and geometry that are different for different racetracks. Geometry of the pinning site affects the depinning current $J_C$ vs. $K_u$ relationship and thus changes the DW velocity in different racetracks. Adding a fixed geometry notch at one end of the racetrack could potentially alleviate the above issue.

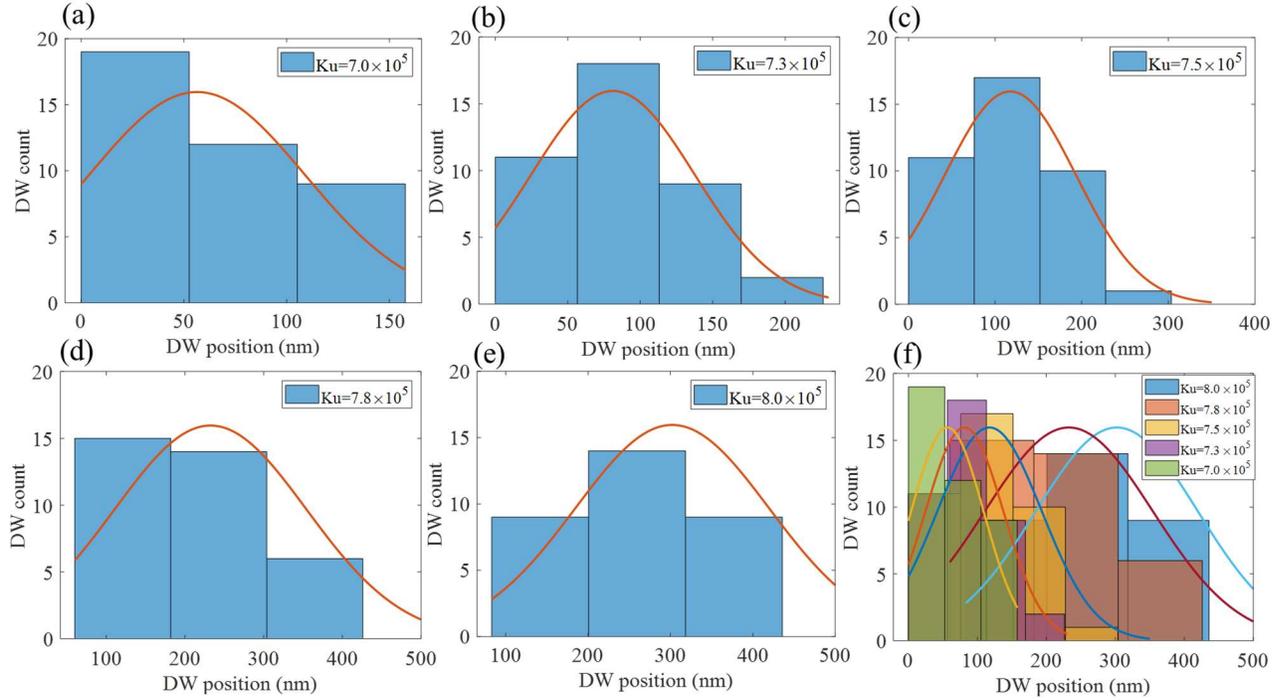

FIG. 3. **a-e.** Equilibrium DW positions for 40 different racetracks at T=0 K for a fixed SOT and different stresses correspond to Ku values of 8.0, 7.8, 7.5, 7.3 and 7.0 ($\times 10^5$) $J/m^3$. For each figure in 3(a-e) a Gaussian distribution plot is overlaid having a mean and standard deviation identical to the data used to create the bins **f.** combined plot of (a-e) shows different mean positions for different $K_u$ values

## 2. Thermal statistics

At room temperature, the thermal perturbation can dislodge the DW. Hence, similar edge roughness cannot offer strong pinning effect. Thus, to achieve better control over DW positions we simulated nanowires with rms edge roughness ~ 6 nm when thermal noise is included. In other words, the racetrack is deliberately designed to be rough at the boundaries to get better performance. We also increase the maximum width to ~100 nm and keep the nanowire length fixed at 500 nm. The fixed clocking SOT current density we use for thermal case is $22 \times 10^{10}$ $A/m^2$ and the SOT and stress application time are kept the same (2 ns). After the withdrawal of SOT and stress, we relax for 10 ns (as we did earlier for the non-thermal case) which gives the DW enough time to reach an equilibrium position. We changed the anisotropy constant, $K_u$ values to 8.0, 7.8, 7.5, 7.3 and 7.0 ($\times 10^5$) $J/m^3$ and ran the simulation 100 times for each Ku value. The equilibrium DW position distribution for one such racetrack is shown in Fig. 4. Here, we also overlay Gaussian distribution with identical mean and standard deviation of the data used to create the bins. The mean positions for different $K_u$ follow the same trend as in non-thermal case where for lower $K_u$ values the mean DW position shifts to the left. Due to the random variation of the DW internal magnetization angle in the presence of thermal noise, upon encountering a potential barrier, the DW could overcome the barrier in some cases but not in other cases. This leads to a distribution.

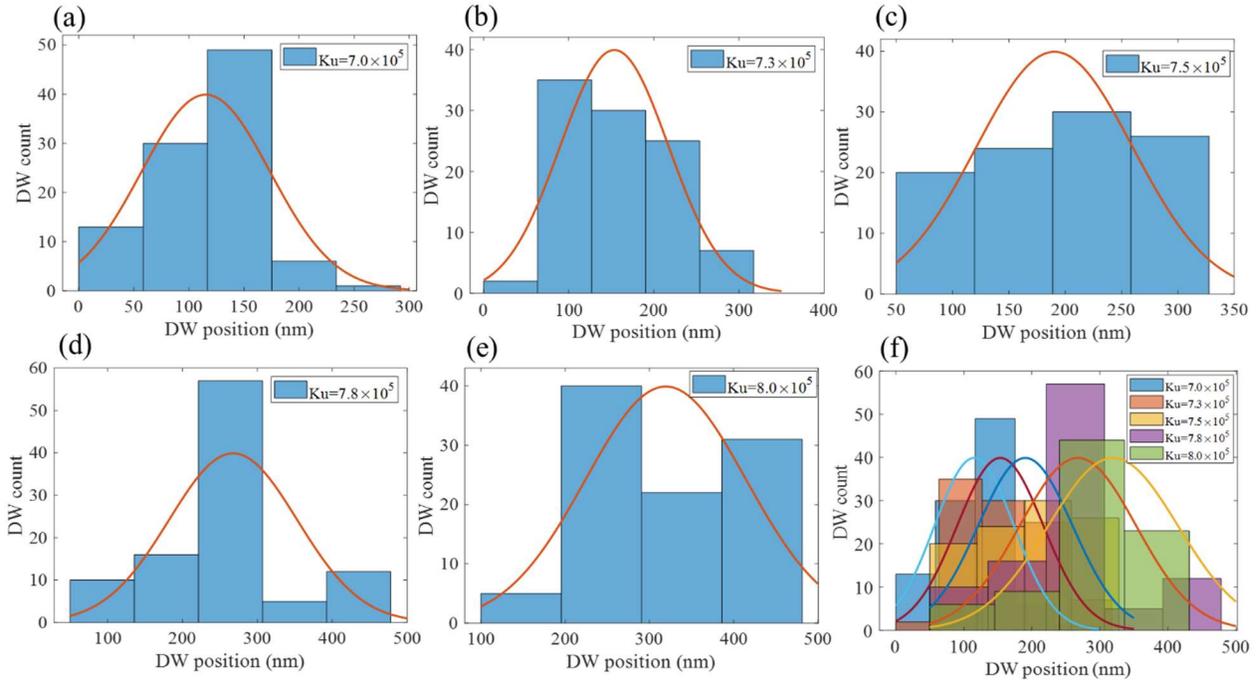

FIG. 4. **a-e.** Equilibrium DW positions for one racetrack at T=300K for a fixed SOT and different stresses correspond to $K_u$ values of 8.0, 7.8, 7.5, 7.3 and 7.0 ($\times 10^5$) $J/m^3$. For each figure in 4(a-e) a Gaussian distribution plot is overlaid having a mean and standard deviation identical to the data used to create the bins **f.** combined plot of (a-e) shows different mean positions for different $K_u$ values.

Although different stress values combined with fixed current pulse can translate the DW into different equilibrium positions, the DW positions are not deterministic as there are considerable overlaps between the positions (Fig. 3 and Fig. 4). The overlaps could be reduced by implementing 3-state (ternary) synaptic weights that can be achieved by choosing only three $K_u$ values. Fig. 5 plots the equilibrium DW positions for a target 3-state synaptic weights where the DW falls within any of the three evenly spaced bins is considered having a resistance value associated with that bin. While 3-state weights reduce the level of overlap between different states, further reduction in overlap may be attainable by selectively positioning notches or protrusions in the racetrack.

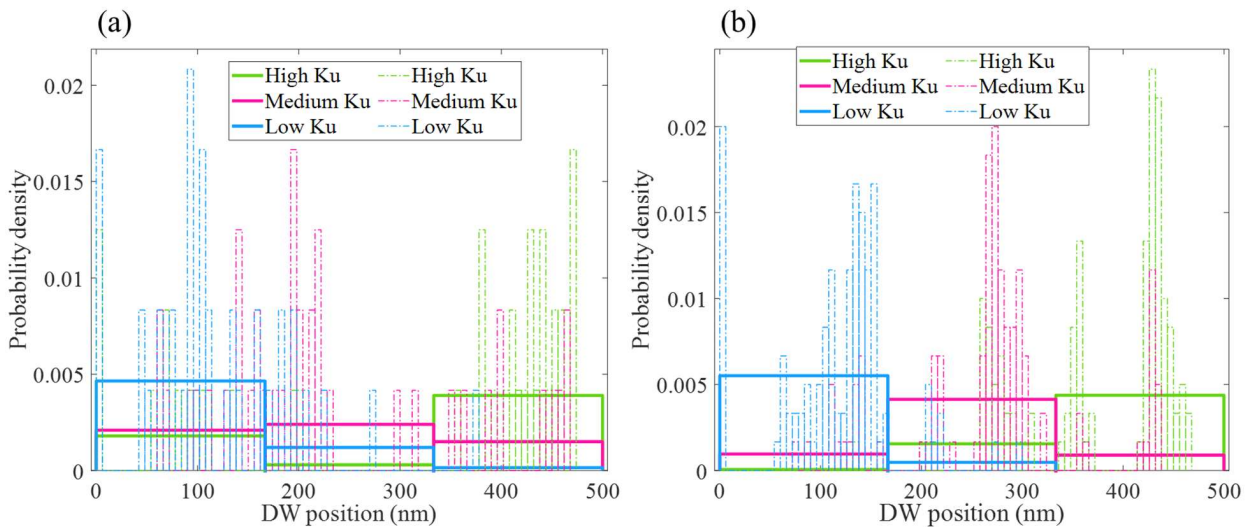

FIG. 5. **a.** Distribution of equilibrium DW position with a fixed SOT and three different stress levels (different Ku) at T=0K for 40 different rough edge (~ 3nm rms) racetracks each having a width and length of 50 nm and 500 nm respectively **b.** Equilibrium DW position distribution with a fixed SOT and three different stress levels at T= 300K for one rough edge (~ 6nm rms) racetrack having a width and length of 100nm and 500 nm respectively. Three equally spaced bins are placed in each of the plots to realize 3-state (ternary) synaptic weight.

While the nanoscale racetrack could be used as a synaptic device, the presence of device to device variation (as in Fig. 3) and intra-device variation (as in Fig. 4) are very evident. Intuitively such variation could be harmful to the functioning of the DW based synaptic device as an inference engine for classification task, as the synaptic weights obtained after software-based training can not be programmed acurately during inference stage. However, recent studies [49] have shown that addressing the device variability during the training stage can achieve high inference accuracies that is very close to baseline accuracy (no device variability is assumed) and the accuracy is highest when the level of noise (because of the device variability) injected during the training is on the same order as the noise of the device used for the inference task.

### 3. Energy dissipation

Energy dissipation in our proposed device depends on charging the piezoelectric layer as well as $I^2R$ loss of the clocking current through the heavy metal layer. To introduce stress, we have to charge the piezoelectric layer. Energy required to charge this capacitive layer is $1/2\, CV^2$, where, V is the voltage applied and C is the capacitance of the piezoelectric layer between the metal contacts.

In our proposed device, the maximum $\Delta$PMA we have used is $0.5\times10^5$ Jm$^{-3}$ and the saturation magnetostriction of CoFe is, $\lambda_s$=250 ppm. Using the above values, the stress σ is calculated to be, $\frac{\Delta PMA}{3/2 \lambda_s}$=133 MPa. For CoFe with Young's Modulus of 200 GPa, the required strain is, $\frac{133\ MPa}{2\times10^5}$ ~$10^{-3}$. Previous study [43] showed that $10^{-3}$ strain is possible in Lead Zirconate Titanate (PZT) piezoelectric with an applied electric field of 3 MVm$^{-1}$. So, if we consider our PZT layer to be 100 nm thick then a voltage of 0.3 V applied at the top electrode pair (see Fig. 1b) can generate the required strain. For a top electrode of length 500 nm (racetrack length 500 nm) and width 100 nm, if the relative permittivity is 3000 then the effective capacitance is calculated to be ~ 13.3 fF. This suggests a $1/2\, CV^2$ loss of ~1.2 fJ considering two top electrodes on both sides of the racetrack.

For our SOT clocking, we assume resistivity of Pt layer to be 100 Ω nm. We assume the Pt layer is 5nm thick, which is greater than the spin diffusion length of ~2 nm [42] and the spin hall angle to be 0.1 [42]. We used current density of $22\times10^{10}$ Am$^{-2}$ through the Pt layer of length 500 nm, width 100 nm and thickness 5nm for a clocking period of 2 ns for the thermal case. So, the $I^2Rt$ loss incurred is ~2.42 fJ. Therefore, our proposed DW based device can reprogram the synaptic weights with a maximum energy dissipation of approximately *3.62 fJ per synapse.*

Energy consumption of the proposed DW-based device is 3.62 fJ per synapse which is much less than previously reported [10] and other DW based devices [11]. Complete CMOS based implementation of a similar device requires a larger number of CMOS which ultimately results in a much larger area density and prohibitive energy cost. Oxide based memristors or phase change memories can also provide multi-level synaptic functionality but their physical mechanism of operation again leads to high energy cost (in the picojoules range) [50].

## 4. Conclusion

In summary, we have proposed an energy efficient strain-controlled synapse where different synaptic weights have been achieved by applying different values of voltage induced stress in conjunction with a fixed clocking SOT current in chiral DW systems with significant DMI. While a uniform change in stress-induced anisotropy cannot move the DW that is pinned in a trap site, it can influence the potential landscape such that the DW in a low PMA racetrack moves faster than in a high PMA one, when being driven by a SOT current. We have showed that five different mean equilibrium DW positions with five different voltage induced stress values is achievable in a 500 nm long and 50 nm wide racetrack with edge roughness ~3 nm. This suggests a 5-state synapse. However, simulations in the presence of thermal noise, demonstrate that even with a higher edge roughness ~6 nm, only a 3-state synapse with three different voltage induced stresses may be achieved in a 500 nm long and 100 nm wide PMA racetrack, if significant overlaps between the distribution of DWs in these different states is to be avoided. Ultimately, this overlap may be reduced by further increasing edge roughness allowing a 5-state synapse at these dimensions in the presence of thermal noise. Recent progress in low precision quantized neural network to achieve near equivalent accuracy to full-precision weight makes such a DW based synapse device specifically attractive as a powerful classification tool for edge devices where the energy requirement is at a premium.

**Acknowledgement:** This work was funded by the National Science Foundation (NSF) Grants ECCS 1954589 and CCF 1815033.

**References:**

[1]. A. Pedram, S. Richardson, M. Horowitz, S. Galal and S. Kvatinsky, Dark Memory and Accelerator-Rich System Optimization in the Dark Silicon Era, IEEE Design & Test 34, 39 (2017)

[2]. M. Suri, O. Bichler, D. Querlioz, O. Cueto, L. Perniola, V. Sousa, D. Vuillaume, C. Gamrat, B. DeSalvo, Phase change memory as synapse for ultra-dense neuromorphic systems: Application to complex visual pattern extraction, 2011 International Electron Devices Meeting (2011)

[3]. G. W. Burr, R. M. Shelby, S. Sidler, C. di Nolfo, J. Jang, Irem Boybat, Rohit S. Shenoy, Pritish Narayanan, Kumar Virwani, Member, Emanuele U. Giacometti, Bülent N. Kurdi and Hyunsang Hwang, Experimental Demonstration and Tolerancing of a Large-Scale Neural Network (165 000 Synapses) Using Phase-Change Memory as the Synaptic Weight Element, IEEE Trans. Electron Devices 62, 3498 (2015)

[4]. I. Boybat, M. L. Gallo, S. R. Nandakumar, T. Moraitis, T. Parnell, T. Tuma, B. Rajendran, Y. Leblebici, A. Sebastian and E. Eleftheriou, Neuromorphic computing with multi-memristive synapses, Nature Communications 9, 2514 (2018)

[5]. S. Yu, Y. Wu, R. Jeyasingh, D. Kuzum, and H.-S. P. Wong, An Electronic Synapse Device Based on Metal Oxide Resistive Switching Memory for Neuromorphic Computation, IEEE Trans. Electron Devices, 58, 2729 (2011)

[6]. P. Yao, H. Wu, B. Gao, S. B. Eryilmaz, X. Huang, W. Zhang, Q. Zhang, N. Deng, L. Shi, H.-S. P. Wong and H. Qian, Face classification using electronic synapses, Nature Communications 8, 15199 (2017)

[7]. A. F. Vincent, J. Larroque, N. Locatelli, N. B. Romdhane, O. Bichler, C. Gamrat, W. S. Zhao, J.-O. Klein, S. G.-Retailleau, and D. Querlioz, Spin-Transfer Torque Magnetic Memory as a Stochastic Memristive Synapse for Neuromorphic Systems, IEEE Transactions on Biomedical Circuits and Systems, 9, 166 (2015)

[8]. S. Ambrogio, P. Narayanan, H. Tsai, R.M. Shelby, I. Boybat, C. di Nolfo, S. Sidler, M. Giordano, M. Bodini, N. C. P. Farinha, B. Killeen, C. Cheng, Y. Jaoudi and G. W. Burr, Equivalent-accuracy accelerated neural-network training using analogue memory, Nature 558, 60 (2018)

[9]. D. Kaushik, U. Singh, U. Sahu, I. Sreedevi, and D. Bhowmik, Comparing domain wall synapse with other non volatile memory devices for on chip learning in analog hardware neural network, AIP Advances 10, 025111 (2020)

[10]. M. A. Azam, D. Bhattacharya, D. Querlioz, C.A. Ross and J. Atulasimha, Voltage control of domain walls in magnetic nanowires for energy-efficient neuromorphic devices, Nanotechnology, 31, 14 (2020)

[11]. S. Lequeux, J. Sampaio, V. Cros, K. Yakushiji, Akio Fukushima, Rie Matsumoto, Hitoshi Kubota, Shinji Yuasa and Julie Grollier, A magnetic synapse: multilevel spin-torque memristor with perpendicular anisotropy, Sci. Rep. 6, 31510 (2016)

[12]. D. M.F. Hartmann, R. A. Duine, M. J. Meijer, H. J.M. Swagten, and R. Lavrijsen, Creep of chiral domain walls, Phys. Rev. B 100, 094417 (2019)

[13]. E. Martinez, S. Emori and G. S. D. Beach, Current-driven domain wall motion along high perpendicular anisotropy multilayers: The role of the Rashba field, the spin Hall effect, and the Dzyaloshinskii-Moriya interaction, Appl. Phys. Lett. 103, 072406 (2013)

[14]. A. V. Khvalkovskiy, V. Cros, D. Apalkov, V. Nikitin, M. Krounbi, K. A. Zvezdin, A. Anane, J. Grollier, and A. Fert, Matching domain-wall configuration and spin-orbit torques for efficient domain-wall motion, Phys. Rev. B 87, 020402(R) (2013)

[15]. D. Bhowmik, M. E. Nowakowski, L. You, O. Lee, D. Keating, M. Wong, J. Bokor and S. Salahuddin Scientific Reports, Deterministic Domain Wall Motion Orthogonal To Current Flow Due To Spin Orbit Torque, 5, 11823 (2015)

[16]. A. Thiaville, Y. Nakatani, J. Miltat, and N. Vernier, Domain wall motion by spin-polarized current: a micromagnetic study, Journal of Applied Physics 95, 7049 (2004)

[17]. P. Chureemart, R. F. L. Evans, and R. W. Chantrell, Dynamics of domain wall driven by spin-transfer torque, Phys. Rev. B 83, 184416 (2011)

[18]. B. Zhang, Y. Xu, W. Zhao, D. Zhu, H. Yang, X. Lin, M. Hehn, G. Malinowski, N. Vernier, D. Ravelosona, and S. Mangin, Domain-wall motion induced by spin transfer torque delivered by helicity-dependent femtosecond laser, Phys. Rev. B 99, 144402 (2019)

[19]. N. Lei, T. Devolder, G. Agnus, P. Aubert, L. Daniel, J.-V. Kim, W. Zhao, T. Trypiniotis, R. P. Cowburn, C. Chappert, D. Ravelosona and P. Lecoeur, Strain-controlled magnetic domain wall propagation in hybrid piezoelectric/ferromagnetic structures, Nature Communications 4, 1378 (2013)

[20]. H.T.Chena and A.K.Soh , Precision electric control of magnetic domain wall motions in a multiferroic bilayer based on strain-mediated magnetoelectric coupling , Materials Research Bulletin 59, 42 (2014)

[21]. Q. Wang, J.Z. Hu, C.Y. Liang, A. Sepulveda and G. Carman, Voltage-induced strain clocking of nanomagnets with perpendicular magnetic anisotropies, Scientific reports, 9, 3639 (2019)

[22]. S. Giordano, Y. Dusch, N. Tiercelin, P. Pernod and V. Preobrazhensky, Combined nanomechanical and nanomagnetic analysis of magnetoelectric memories, Physical Review B 85, 155321 (2012)

[23]. K. Roy, S. Bandyopadhyay, and J. Atulasimha, Hybrid spintronics and straintronics: A magnetic technology for ultra low energy computing and signal processing, Appl. Phys. Lett. 99, 063108 (2011)

[24]. J. Atulasimha, and S. Bandyopadhyay, Bennett clocking of nanomagnetic logic using multiferroic single-domain nanomagnets, Appl. Phys. Lett. 97, 173105 (2010)

[25]. K. Roy, S. Bandyopadhyay and J. Atulasimha, Binary switching in a 'symmetric' potential landscape, Scientific Reports **3**, 3038 (2013)

[26]. N. D'Souza, M. S. Fashami, S. Bandyopadhyay and J. Atulasimha, Experimental Clocking of Nanomagnets with Strain for Ultralow Power Boolean Logic, Nano Lett. 16, 1069 (2016)

[27]. A. K Biswas, H. Ahmad, J. Atulasimha and S. Bandyopadhyay, Experimental Demonstration of Complete 180° Reversal of Magnetization in Isolated Co Nanomagnets on a PMN–PT Substrate with Voltage Generated Strain, Nano Lett. 17, 3478 (2017)

[28]. A. W. Rushforth, R. R. Robinson and J Zemen, Deterministic magnetic domain wall motion induced by pulsed anisotropy energy, J. Phys. D: Appl. Phys. 53 ,164001 (2020)

[29]. T. Mathurin, S. Giordano, Y. Dusch, N. Tiercelin, P. Pernod and V. Preobrazhensky, Stress-mediated magnetoelectric control of ferromagnetic domain wall position in multiferroic heterostructures, Appl. Phys. Lett. 108, 082401 (2016)

[30]. V. Uhl´ıˇr, S. Pizzini, N. Rougemaille, J. Novotn´y, V. Cros, E. Jim´enez, G. Faini, L. Heyne, F. Sirotti, C. Tieg, A. Bendounan, F. Maccherozzi, R. Belkhou, J. Grollier, A. Anane, and J. Vogel, Current-induced motion and pinning of domain walls in spin-valve nanowires studied by XMCD-PEEM, Phys. Rev. B 81, 224418 (2010)

[31]. X. Jiang, L. Thomas, R. Moriya, M. Hayashi, B. Bergman, C. Rettner and S. S.P. Parkin, Enhanced stochasticity of domain wall motion in magnetic racetracks due to dynamic pinning, Nature Communications 1, 25 (2010)

[32]. J. P. Attan´e, D. Ravelosona, A.Marty, Y. Samson, and C. Chappert, Thermally Activated Depinning of a Narrow Domain Wall from a Single Defect, Phys. Rev. Lett. 96, 147204 (2006)

[33]. R. Lewis, D. Petit, L. Thevenard, A. V. Jausovec, L. O'Brien, D. E. Read, and R. P. Cowburn, Magnetic domain wall pinning by a curved conduit, Appl. Phys. Lett. 95, 152505 (2009)

[34]. D. Petit, A.-V. Jausovec, D. Read, and R. P. Cowburn, Domain wall pinning and potential landscapes created by constrictions and protrusions in ferromagnetic nanowires, J. Appl. Phys. 103, 114307 (2008)

[35]. M. Albert, M. Franchin, T. Fischbacher, G. Meier and H. Fangohr, Domain wall motion in perpendicular anisotropy nanowires with edge roughness, Journal of Physics: Condensed Matter, 24, 2 (2011)

[36]. S. Dutta, S. A. Siddiqui, J. A. Currivan-Incorvia, C. A. Ross, and M. A. Baldo, The Spatial Resolution Limit for an Individual Domain Wall in Magnetic Nanowires, Nano Lett. 17, 5869 (2017)

[37]. I. Hubara, M. Courbariaux, D. Soudry, R. El-Yaniv, Y. Bengio, Quantized Neural Networks: Training Neural Networks with Low Precision Weights and Activations, The Journal of Machine Learning Research 18, 1-30 (2017)

[38]. B. Jacob, S. Kligys, B. Chen, M. Zhu, M. Tang, A. Howard, H. Adam, D. Kalenichenko, Quantization and Training of Neural Networks for Efficient Integer-Arithmetic-Only Inference, arXiv:1712.05877

[39]. F. Li, B. Zhang, B. Liu, Ternary weight networks, arXiv:1605.04711

[40]. S. Emori, U. Bauer, Sung-Min Ahn, E. Martinez and G. S. D. Beach, Current-driven dynamics of chiral ferromagnetic domain walls, Nat. Mater. 12, 611 (2013)

[41]. A. Vansteenkiste, J. Leliaert, M. Dvornik, M. Helsen, F. Garcia-Sanchez, and B. V. Waeyenberge, The design and verification of MuMax3, AIP Advances 4, 107133 (2014)

[42]. L. Liu, R. A. Buhrman, D. C. Ralph, Review and Analysis of Measurements of the Spin Hall Effect in Platinum, arXiv:1111.3702 (2011)

[43]. J. Cui, J. L. Hockel, P. K. Nordeen, D. M. Pisani, Cheng-yen Liang, Gr. P. Carman, and C. S. Lynch Appl. Phys. Lett., A method to control magnetism in individual strain-mediated magnetoelectric islands, 103, 232905 (2013)

[44]. C. Bilzera, T. Devolder, Joo-Von Kim, G. Counil, and C. Chappert, Study of the dynamic magnetic properties of soft CoFeB films, J. Appl. Phys. 100 053903 (2006)

[45]. M. Belmeguenai, M. S. Gabor, Y. Roussigné, A. Stashkevich, S. M. Ch´erif, F. Zighem, and C. Tiusan, Brillouin light scattering investigation of the thickness dependence of Dzyaloshinskii-Moriya interaction in $Co_{0.5}Fe_{0.5}$ ultrathin films, Physical Review B 93, 174407 (2016)

[46]. D. Hunter, W. Osborn, K. Wang, N. Kazantseva, J. H.-Simpers, R. Suchoski, R. Takahashi, M. L. Young, A. Mehta, L. A. Bendersky, S. E. Lofland, M. Wuttig and I. Takeuchi, Giant magnetostriction in annealed $Co_{1-x}Fe_x$ thin-films, Nature Communications 2, 518 (2011)

[47]. A. Thiaville, S. Rohart, É. Jué, V. Cros and A. Fert, Dynamics of Dzyaloshinskii domain walls in ultrathin magnetic films, EPL (Europhysics Letters) 100, 5 (2012)

[48]. J. Torrejon, E. Martinez and M. Hayashi, Tunable inertia of chiral magnetic domain walls, Nature Communications **7**, 13533 (2016)

[49]. V. Joshi, M. Le Gallo, S. Haefeli, I. Boybat, S. R. Nandakumar, C. Piveteau, M. Dazzi, B. Rajendran, A. Sebastian & E. Eleftheriou, Accurate deep neural network inference using computational phase-change memory, Nature Communications 11, 2473 (2020)

[50]. M. Prezioso, F. M. Bayat, B. D. Hoskins, G. C. Adam, K. K. Likharev, D. B. Strukov, Training and operation of an integrated neuromorphic network based on metal-oxide memristors, Nature 521, 61 (2015)